\documentstyle[11pt]{article}
\newcommand{\AmS}{{\protect\the\textfont2
  A\kern-.1667em\lower.5ex\hbox{M}\kern-.125emS}}
  \def\d{\delta} \def\D{\Delta} 
\def\f{\phi} \def\vf{\varphi} \def\la{\lambda}  
 \def\p{\pi} \def\x{\chi}  \def\r#1{$^{[#1]}$}
\def\E{\mbox{e}^+\mbox{e}^-}

\def\cent{\centerline} \def\pa{\parindent} 
  \def\vs{\vskip} \def\ni{\noindent}
\def\pa{\parindent} \def\ej{\vfill\eject} \def\hf{\hfill} 
  
\def\ifmath#1{\relax\ifmmode #1\else $#1$\fi}%
 
\def\ra{\ifmath{{\mathrm{a}}}} 
\def\rb{\ifmath{{\mathrm{b}}}} \def\rc{\ifmath{{\mathrm{c}}}}
\def\rd{\ifmath{{\mathrm{d}}}}

 \def\rT{\ifmath{{\mathrm{T}}}}

\newcommand{\beqa}{\begin{eqnarray}} \newcommand{\eeqa}{\end{eqnarray}  }
\newcommand{\beqan}{\begin{eqnarray*}} \newcommand{\eeqan}{\end{eqnarray*}}
\newcommand{\beq}{\begin{equation}} \newcommand{\eeq}{\end{equation}  }
\begin{document}

\hf HEN 406

\hf HZPP-9805

\hf May 5, 1998

\vs 5mm
\centerline{\Large\bf Self-affine scaling from non-integer phase-space
partition}

\centerline{\Large\bf in $\pi^+$p and K$^+$p Collisions at 250 GeV/$c$}
\vskip0.5cm

\cent{EHS/NA22 Collaboration}

\vs  5mm
 
{\pa=0pt
N.M. Agababyan$^{8}$, 
M.R. Atayan$^{8,\ra}$, 
Chen Gang$^{7,\rb}$, 
E.A. De Wolf$^{1,\rc}$, 
K. Dziunikowska$^{2,\rd}$,\\ 
A.M.F. Endler$^{5}$,
Gao Yanmin$^{7}$, 
Z.Sh. Garutchava$^{6}$, 
H.R. Gulkanyan$^{8,\ra}$,
R.Sh. Hakobyan$^{8,\ra}$,
D. Kisielewska$^{2,\rd}$,
W. Kittel$^{4}$, 
Liu Lianshou$^{7}$,
S.S. Mehrabyan$^{8,\ra}$,
Z.V. Metreveli$^{6}$, 
K. Olkiewicz$^{2,\rd}$, 
F.K. Rizatdinova$^{3}$, 
E.K. Shabalina$^{3}$, 
L.N. Smirnova$^{3}$,
M.D. Tabidze$^{6}$, 
L.A. Tikhonova$^{3}$,\\
A.V. Tkabladze$^{6}$,
A.G. Tomaradze$^{6}$,
F. Verbeure$^{1}$,
Wu Yuanfang$^{7}$,
S.A. Zotkin$^{3}$ 

\begin{itemize}
\itemsep=-2mm
\item[$^{1}$] Department of Physics, Universitaire Instelling Antwerpen,
B-2610 Wilrijk, Belgium
\item[$^{2}$] Institute of Physics and Nuclear Techniques of Academy of
Mining and Metallurgy and Institute of Nuclear Physics, PL-30055 Krakow,
Poland
\item[$^{3}$] Nuclear Physics Institute, Moscow State University, 
RU-119899 Moscow, Russia
\item[$^{4}$] High Energy Physics Institute Nijmegen (HEFIN), University of 
Nijmegen/NIKHEF, NL-6525 ED Nijmegen, The Netherlands
\item[$^{5}$] Centro Brasileiro de Pesquisas Fisicas, BR-22290 
Rio de Janeiro, Brazil
\item[$^{6}$] Institute for High Energy Physics of Tbilisi State 
University, GE-380086 Tbilisi, Georgia
\item[$^{7}$] Institute of Particle Physics, Hua-Zhong Normal 
University, Wuhan 430070, China
\item[$^{8}$] Institute of Physics, AM-375036 Yerevan, Armenia
\end{itemize}

\vskip0.5cm
{\leftskip=1.cm\rightskip=1.cm
{\bf Abstract} \\
A factorial-moment analysis with real (integer and non-integer) phase 
space partition is applied to $\p^+$p and K$^+$p collisions at 250 GeV/$c$.
Clear evidence is shown for self-affine rather than self-similar power-law 
scaling in multiparticle production. The three-dimensional self-affine
second-order scaling exponent is determined to be 0.061$\pm$0.010. 
\par}
 
\vfill

\hrule width3truecm
\vs 2mm
{\small\ni
$^a$ Supported by the Government of the Republic of Armenia, contractnumber
94-496\\
$^b$ Permanent address: Department of Physics, Jingzhou Teacher's College,
Hubei 434100, China\\
$^c$ Onderzoeksdirecteur NFWO, Belgium\\
$^d$ Supported by the Polish State Committee for Scientific Research
}\par}
 
\ej

\section{Introduction}
\vs-0.2truecm

Classical non-Abelian theories show non-linear behavior \r{1} and the 
non-linear development of a classical non-Abelian system can be determined
numerically \r{2}. Although extension to a quantum theory as QCD is far 
from trivial, non-linearity is inherent also to QCD shower development.\r{3}

The first experimental evidence for non-linear behavior in high-energy 
multiparticle production came from multiplicity fluctuations in a JACEE event 
recorded in 1983 \r{4,5}. In this event, the total multiplicity is about one 
thousand and the fluctuations in small rapidity bins are 2 times average. In 
1987, NA22 \r{6} found an event in which the fluctuation in a small rapidity 
bin is as high as 60 times average. 

To be able to decide whether these fluctuations are dynamical, i.e. larger
than expected from Poisson noise, Bia\l as and Peschanski \r{6} suggested 
to use factorial moments (FM), defined as 
\begin{equation}
F_q(\delta)=\frac{1}{M} \sum \limits_{m=1} ^M \frac {\langle  n_m (n_m-1) 
\cdots (n_m-q+1) \rangle} {\langle n_m \rangle ^q} .
\end{equation}   
In the above equation, $M$ is the partition number of a fixed phase space
region $\Delta$ under consideration, $\delta=\Delta/M$ is the size of 
a sub-cell, $n_m$ is the number of (charged) particles falling into the 
$m$th sub-cell. The authors show, if the power-law scaling 
\beq
F_q(\delta)\propto \delta ^{-\phi_q} 
\eeq 
holds when $\delta \to 0$, then dynamic self-similar fluctuations are present 
in multiparticle production. 

In the following, the behavior of $F_q$ in ever smaller phase-space 
cells $\delta$ was studied in almost all high-energy experiments on 
lepton-lepton, lepton-nucleon, hadron-hadron and nucleus-nucleus collisions 
and approximate scaling was established (for recent reviews see \r{7}). 
It has been pointed out \r{8}, however, that the anisotropy of phase 
space \r{9} has to be taken into account and that the multiparticle final 
state in a high energy hadron-hadron collision may be 'self-affine' \r{10} 
rather than self-similar. 

In this note we apply a self-affine analysis to the NA22 data on
$\p^+$p and K$^+$p collisions at 250 GeV/$c$, where, in addition, the
phase-space partition $M$ of (1) is generalized to real (i.e. also
non-integer) values. This method reduces the $\x^2$/NDF values by a
factor of 2 with respect to the limitation to integer values and grants
clear evidence for the presence of dynamical self-affine fluctuations
in these collisions. The three-dimensional self-affine scaling exponent
is determined to be $\phi^{3D}_2=0.061\pm0.010$.

\section{The method}
A self-affine transformation in the three phase-space variables denoted 
$p_a, p_b, p_c$ is defined as $\d p_a\to \d p_a/\la_a$,
$\d p_b\to \d p_a/\la_b$, $\d p_c \to \d p_c/\la_c$, with 
shrinking ratios $\la_a$, $\la_b$ and $\la_c$, respectively. 
The anisotropy (self-affinity) of a dynamical fluctuation can then be 
characterized by the so-called roughness or Hurst exponents \r{10} 
\begin{equation}
H_{ij}={\ln \la_i\over \ln \la_j},\ \ \ \ \ \ \  (i,j=a,b\quad \mbox{or} \quad
 a,c\quad
\mbox{or} \quad b,c),  
\end{equation}   
with
\vs -10mm 
\begin{equation}
\la_i \le \la_j, \quad \quad 0\le H_{ij}\le 1. 
\end{equation}   
These exponents can be obtained \r{11} from the experimental $\d$ dependence
observed in the second-order factorial moments in the three corresponding 
variables. If self-affine fluctuations of multiplicity exist in multiparticle 
production, exact scaling, i.e. a straight line in ln$F_q$ versus ln$M$, 
should be observed if and only if the $\la_i$ are allowed to differ from one 
another.

The above prediction was checked on our data \r{12} and on 400 GeV/$c$ pp 
data.\r{13} The Hurst exponents for longitudinal-transverse directions were 
determined to be 0.474$\pm$0.056 in the ($y,\vf$) plane and 0.477$\pm$0.057
in the ($y,p_\rT$) plane for 250 GeV/$c$ $\pi^+$p and K$^+$p collisions and 
0.74$\pm$0.07 in the ($\eta,\vf$) plane for 400 GeV/$c$ pp collision. A 
self-affine higher-dimensional analysis\footnote{In the case of NA27, due to 
lack of momentum measurement, the self-affine analysis was 2D instead of 3D.} 
was performed in both cases and the results are confirmative.

However, the original analysis was limited by the small number of combinations
of (integer) partitions $M_i$ allowed by condition (3). Therefore, the method 
(and necessary correction procedure) has recently been generalized to 
non-integer partitions $M$.\r{14} In this letter the new method and correction
procedure are applied to obtain a precise answer on the question of 
self-affinity in our data.

\subsection{Non-integer FM analysis\r{14}}
\vs-0.2truecm

To be definite, let us consider a one-dimensional analysis in rapidity $y$.
In the ideal case, the factorial moments $F_q(\delta y)$ depend on the bin 
width $\delta y=\D y/M$, but not on the position of the bin on the rapidity
axis. If that is the case, the result of averaging over all $M$ bins as
in (1) is equal to that of averaging over $N$ bins with $N\leq M$.
This means that ideally one has $F_2(M) =F_2(N,M)$, where
\begin{equation}
 F_2(N,M)=\frac{1}{N} \sum \limits_{m=1} ^N
\frac {\langle  n_{m}  (n_{m}-1) \rangle} {\langle n_m \rangle ^2},
 \qquad (N \leq M, \d y = \D y/M)\ . \end{equation}       
This equation can be used as the definition of FM for any real (integer or 
non-integer) value of $M$ \r{15}, provided that the number $N$ of bins used 
for averaging is taken to be
\begin{equation}
              N=M-a, \qquad (0\leq a<1)  .
\end{equation}       

However, even in the central region the rapidity distribution is not flat. 
The shape of this distribution influences the scaling behavior of the FM. 
Therefore, the cumulant variable 
\begin{equation}
  x(y)= \frac{\int_{y_a}^y \rho(y') dy'} {\int_{y_a}^{y_b} \rho(y') dy'} 
\end{equation}      
was introduced,\r{16} which has a flat distribution by definition.
An additional correction factor has to be introduced again for the FM 
analysis with non-integer partition.

To see this, let $\Delta$ denote the phase-space region in consideration, 
$\delta_m$ the $m$th bin, $\rho_1(y_1)$ and $\rho_2(y_1,y_2)$ the one- and 
two-particle distribution functions, respectively. Then we have
$$ \langle n_m\rangle = \int_{\delta_m}\rho_1(y)dy
  = {\langle n \rangle \over \langle n(n-1)\rangle}
 \int_{\Delta} dy_2\int_{\delta_m} dy_1 \rho_2(y_1,y_2); \eqno(8a)$$
$$\langle n_m(n_m-1)\rangle= \int_{\delta_m}dy_2 \int_{\delta_m} 
   dy_1\rho_2(y_1,y_2). \eqno(8b)$$

After transforming to the cumulant variable, $\langle n_m\rangle$ becomes 
a constant, independent of $m$. However, comparing the two above 
equations, it can be seen that due to the difference in the integration 
region over $y_2,\langle n_m(n_m-1)\rangle$ is in general not constant
even though $\langle n_m\rangle$ is. This was 
experimentally verified on our data.

Note that in the definition of $F_q$ in (1) a horizontal average is taken. 
When the partition number $M$ is an integer, the horizontal average is 
over the full region $\Delta$. The variation of 
$\langle n_m(n_m-1)\rangle/\langle n_m\rangle^2$ 
is thus smeared out, and no correction is needed. On the contrary, when $M$ 
is non-integer, the horizontal average is performed over only part of the 
region and the influence of the variation of 
$\langle n_m(n_m-1)\rangle/\langle n_m\rangle^2$ becomes essential. 

\subsection{Correction factor for the 
$\langle n_m(n_m-1)\rangle/\langle n_m\rangle^2$ distribution}
\vs-0.2truecm

From the  definition of $F_2(N,M)$ in (5) it can be seen that only $N$ bins 
are included in the horizontal average when $M$ is non-integer 
($M=N+a,\  0< a<1)$. Consequently, only a fraction $r=N/M$ of the full
region $\Delta$ is taken into account. We minimize the influence 
of this by introducing a correction factor $R(r)$ and define \r{14}
$$ F_2(M)=\frac{1} {R(r)} \left(\frac{1}{N} \sum \limits_{m=1} ^N
\frac {\langle  n_{m}  (n_{m}-1) \rangle} {\langle n_m \rangle ^2}\right).
\eqno(9) $$

In order to extract $R(r)=R(N/M)$ from the experimental data, let us go 
back to integer $M$ and calculate $F_2$ averaging only over $N$ of the $M$ 
bins ($N\leq M$). The result is, in general, a function of both $N$ and $M$. 
The correction matrix 
 $$ C(N,M)= \frac {\frac{1}{N} \sum \limits_{m=1} ^N
    \langle  n_{m}  (n_{m}-1) \rangle/\langle n_m\rangle^2}
 {\frac{1}{M} \sum \limits_{m=1} ^M
    \langle  n_{m}  (n_{m}-1) \rangle/\langle n_m\rangle^2} \ , \ \ \ \ 
 N=1,2,\dots,M.
                          \eqno(10)  $$
is shown in Fig.~1. as a function of $N/M$ for $M=3,4,\dots,40$. Points 
for different $M$ lie in a narrow band (mind the scale), so that the 
correction factors $R_y(r)$, $R_{p_\rT}(r)$ and $R_\vf(r)$ can be obtained 
from an interpolation of $C(N,M)$. 

It turns out, however, that the result is very sensitive to the interpolation 
function. An inappropriate choice of the interpolation function, even inside 
the narrow $C(N,M)$ band, will be either insufficient in eliminating
the ``sawteeth'' observed in the uncorrected ln$F_2$ versus ln$M$ plot or 
it will over-correct. It is found \r{14} that when the ``sawteeth'' 
lie above the smooth curve of integer $M$, as in 
the case of $y$ and $p_\rT$, the upper boundary of the $C(N,M)$ band has to 
be used for the interpolation. When the ``sawteeth'' reach from above the 
smooth curve of integer $M$ to below, as in the case of $\vf$, the middle 
of the $C(N,M)$ band has to be used. The interpolation functions used for 
the three cases are shown as dotted lines in Fig.~1. 

Having obtained the correction factor for one-dimensional FM's with
non-integer $M$, the correction of higher-dimensional FM's can be obtained 
from a straight-forward generalisation \r{14} paying special attention to 
the overlap regions $(M_1-N_1)(M_2-N_2)$ and $(M_1-N_1)(M_2-N_2)(M_3-N_3)$ 
in the two- and three-dimensional analysis, respectively.

\section{Results}
\vs-0.2truecm

The details of the EHS spectrometer can be found in \r{17}, those of the 
trigger and data analysis in \r{18}. The acceptance criteria and data 
samples are those already used in \r{12}. 

\subsection{One-dimensional analysis and Hurst parameters}

In Fig.~2 are shown the results of a one-dimensional analysis of ln$F_2$ 
versus ln$M$. The first column reproduces the results obtained earlier for 
integer $M$ \r{12}. The second column gives the results for real $M$ as 
defined in (5). It can be seen that the points for non-integer $M$ depart 
in a ``sawtooth pattern'' from the curve determined by the points with 
integer $M$, especially in the cases of $y$ and $p_\rT$. After correction 
by $R_y$, $R_{p_\rT}$ and $R_\vf$ obtained from the interpolation of the 
correction matrix $C(N,M)$ above, the results (shown  in the third column 
of Fig.~2) become smooth.

\subsection{Two-dimensional analysis}

The plots for 2-dimensional self-affine FM's are presented in Fig.~3
for $H_{p_\rT\vf}=1$ and $H_{y p_\rT}=H_{y \vf}=0.475$. 
The same 1-dimensional correction factors $R_y$, $R_{p_\rT}$ and
$R_\vf$ are used together with the geometrical factors taking care of
the overlap regions \r{14}. The corrected results for ($y,p_\rT$) and 
($y,\vf$) are satisfactory. For ($p_\rT,\vf$) $F_2$ is improved considerably 
with respect to the original real (integer and non-integer) $M_y$ plot shown 
in the middle column of Fig.~2.

\subsection{Three-dimensional analysis}

The final 3-D results are presented in Fig.~4. The left figure is the one
with integer $M$ obtained before \r{13}. The right figure is the corrected
result of the self-affine analysis with real $M$ together with a linear fit 
$$ \ln F_2=A+B\ln M_y\ . \eqno(11) $$ 
In order to minimize the influence of momentum conservation \r{19}, the fit 
starts from $M_y=2$. For the same reason, the non-integer $M$ points are 
given for $M_y>2$, only.

It has to be noted that integer M values could be used in our previous
analysis only because the Hurst exponents had values close to 0.5 and
1.0, respectively. This was rather accidental, however, and a generalization
to real (integer and non-interger) partition \r{15} is necessary to be able 
to evaluate other experiments (e.g.\r{13}). A comparison of integer-M and 
real-M results in table 1 shows that the slope $B$ itself is unchanged but 
the error tends to decrease. Furthermore, it can be seen from table 1 that 
the ratios $\chi^2$/NDF are decreased as compared to those obtained from 
integer $M$ and at the same time the number of degrees of freedom (NDF) 
increases largely.

Note that the slope $B$ of (11) is related to $\f_2$ of (2) via
$$M_{3D}=M_y M_{p_\rT} M_\vf = M_y^{1+\frac{1}{H_{y p_\rT}}+
\frac{1}{H_{y\vf}}}$$
as
$$\f_2 = B\left/\left(1+\frac{1}{H_{yp_\rT}}+\frac{1}{H_{y\vf}}\right)\right.
\ . \eqno(12)$$
From the $B$ value for weighted real $M$ in table 1, we get as the final 
result for the scaling exponent
$$\f^{3D}_2 = 0.061\pm 0.010 \ . \eqno(13)$$

\section{Conclusions}
\vs-0.2truecm

In this paper we presented the results of a self-affine analysis of
experimental data on factorial moments generalized to real (integer and 
non-integer) partition $M$. Correction factors were introduced to minimize 
the influence of the variation of $\langle n_m(n_m-1)\rangle$. The corrected 
results for non-integer $M$ lie on smooth lines interpolating between the 
integer-$M$ points obtained before.

The results of a three-dimensional self-affine analysis are well fitted by a 
power law, giving a more general and more precise check of the self-affine 
power-law scaling of the data than the previous analysis based on integer 
partition only. The observed behavior is consistent with the fact that the 
longitudinal direction is privileged over the transverse directions in 
hadron-hadron collisions. It would be important to use this analysis in the 
study of QCD parton-shower development in $\E $ collisions.

\section*{Acknowledgements}
We are grateful to the III. Physikalisches Institut B, RWTH Aachen, Germany, 
the DESY-Institut f\"ur Hochenergiephysik, Berlin-Zeuthen, Germany, the 
Institute for High Energy Physics, Protvino, Russia, the Department of 
High Energy Physics, Helsinki University, Finland, and the University of 
Warsaw and Institute of Nuclear Problems, Poland for early contributions 
to this experiment.  This work is part of the research program of the 
``Stichting voor Fundamenteel Onderzoek der Materie (FOM)", which is 
financially supported by the ``Nederlandse Organisatie voor Wetenschappelijk 
Onderzoek (NWO)". We further thank NWO for support of this project within 
the program for subsistence to the former Soviet Union (07-13-038), as well 
as the National Commission of Science and Technology of China and the Royal 
Academy of Science of the Netherlands for support within the program Joint 
Research between China and the Netherlands under project number 97CDP004.
The support from NSFC and OYTF of NCEC is appreciated.

\ej

\hskip2.2cm Table 1. \ \ The parameter values obtained from a fit of the
3D-data by (11)
\vskip0.5cm

\hskip2.0cm
\begin{tabular}{|r|c|c|c|}\hline
\null\hfill Method \hfill\null & $A$ & $B$ & $\chi^2/$NDF \\ \hline
Without bin-size correlation & & & \\
          weighted integer-M & -0.04$\pm$0.03 & 0.32$\pm$0.03 & 7/4 \\
             weighted real-M & -0.01$\pm$0.02 & 0.32$\pm$0.02 & 19/20 \\ \hline
Without bin-size correlation & & & \\
        unweighted integer-M & -0.08$\pm$0.02 & 0.33$\pm$0.03 & 12/4 \\ 
           unweighted real-M & -0.05$\pm$0.02 & 0.33$\pm$0.02 & 30/20 \\ \hline
With bin-size correlation    & & & \\
        unweighted integer-M & -0.08$\pm$0.02 & 0.34$\pm$0.02 & 14/4 \\
           unweighted real-M & -0.05$\pm$0.02 & 0.34$\pm$0.02 & 51/20 \\ \hline
\end{tabular}

\ej

\noindent{\bf\large\bf Figure captions}
\vskip0.1cm

\noindent 
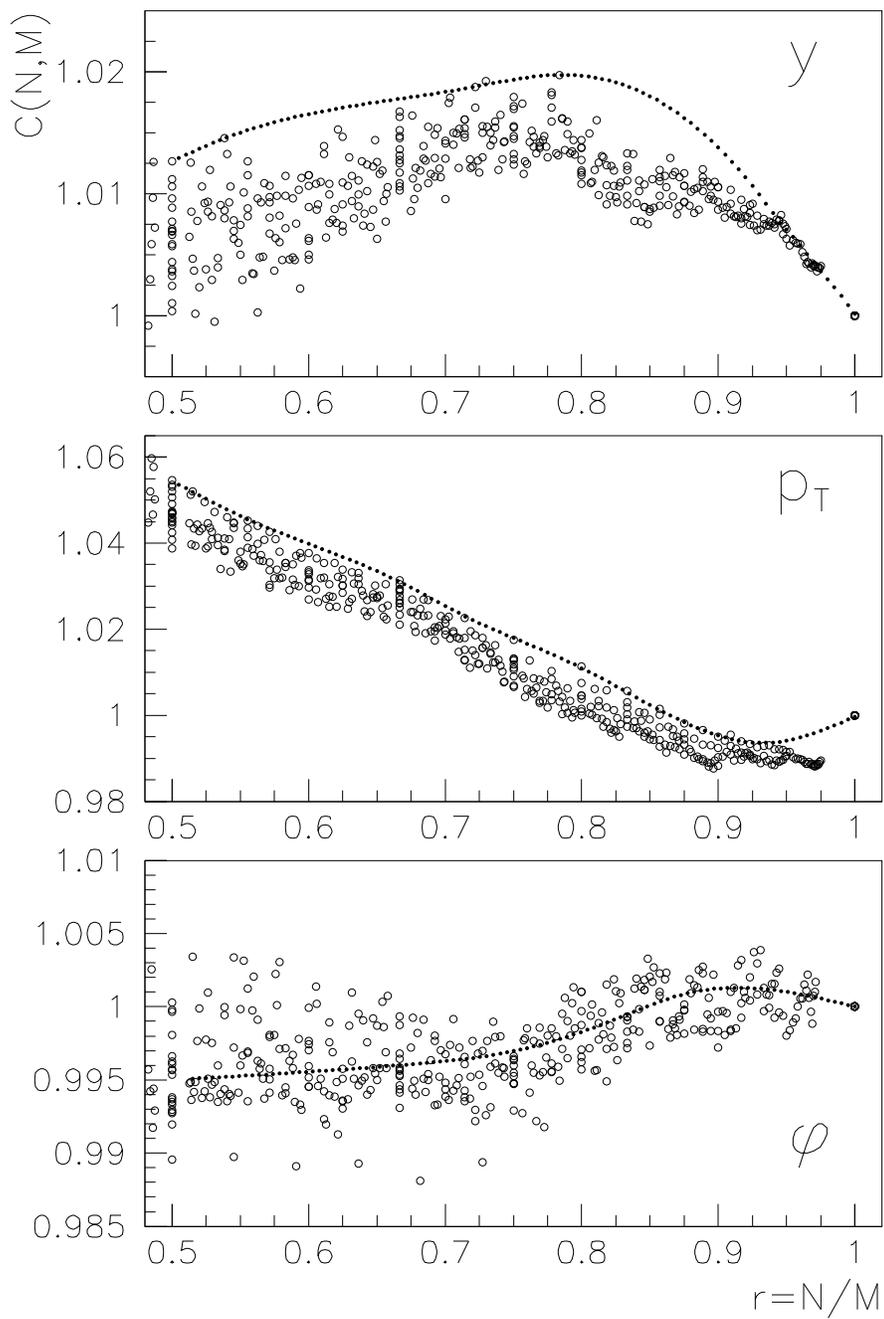
Fig.~1 \ The correction matrix $C(N,M)$ as function of $N/M$ for 
$M=3$ -- $40$ (for the curves see text).

\noindent 
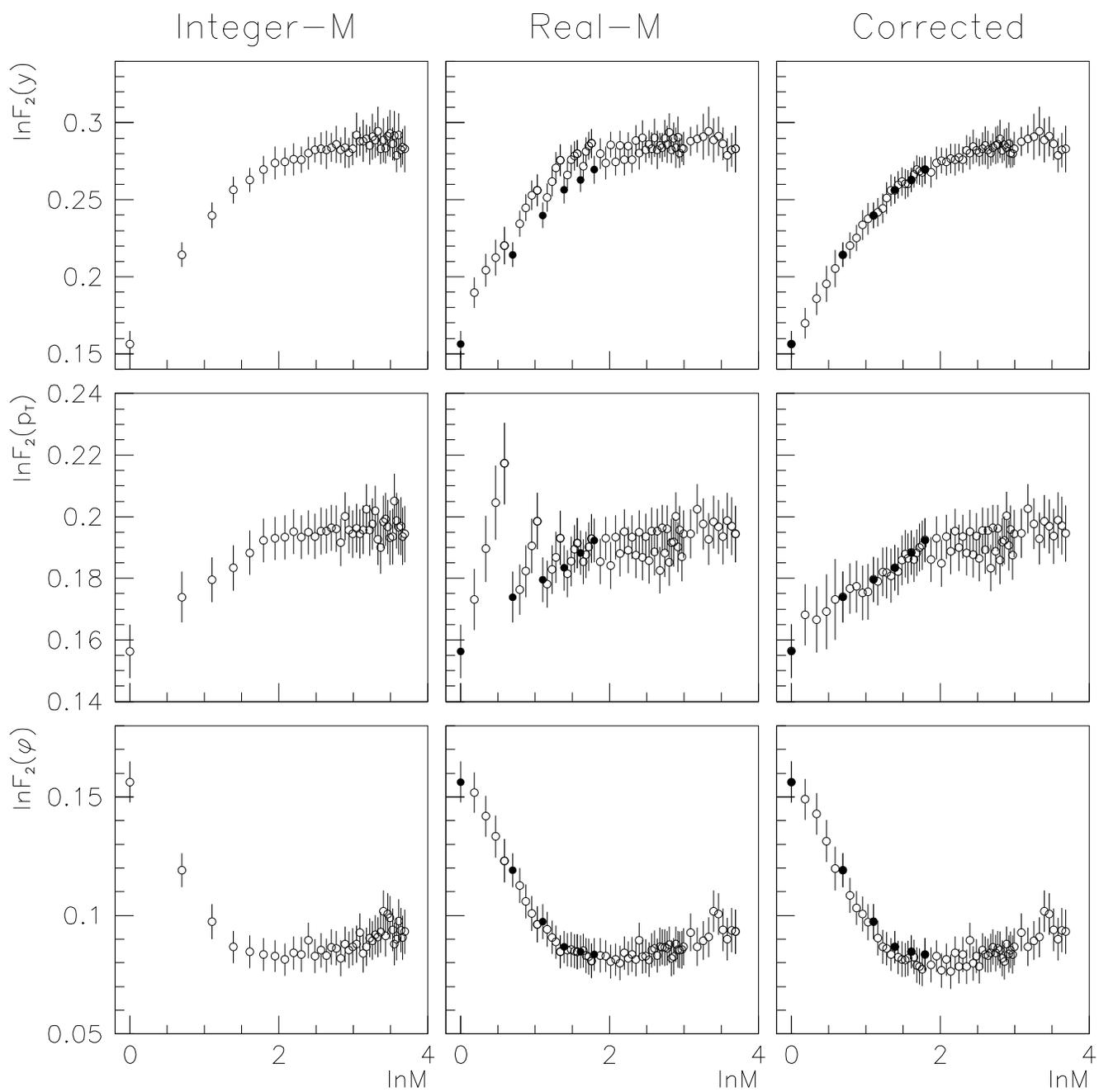
Fig.~2 \ The one-dimensional plots of ln$F_2$ versus ln$M$. The first 
column are the previous results for integer $M$. The second column are 
the results of real $M$ as defined in (5). The third column are the 
results after correction. For easier comparison, the first six integer-M 
points are indicated as full circles in columns two and three.

\noindent 
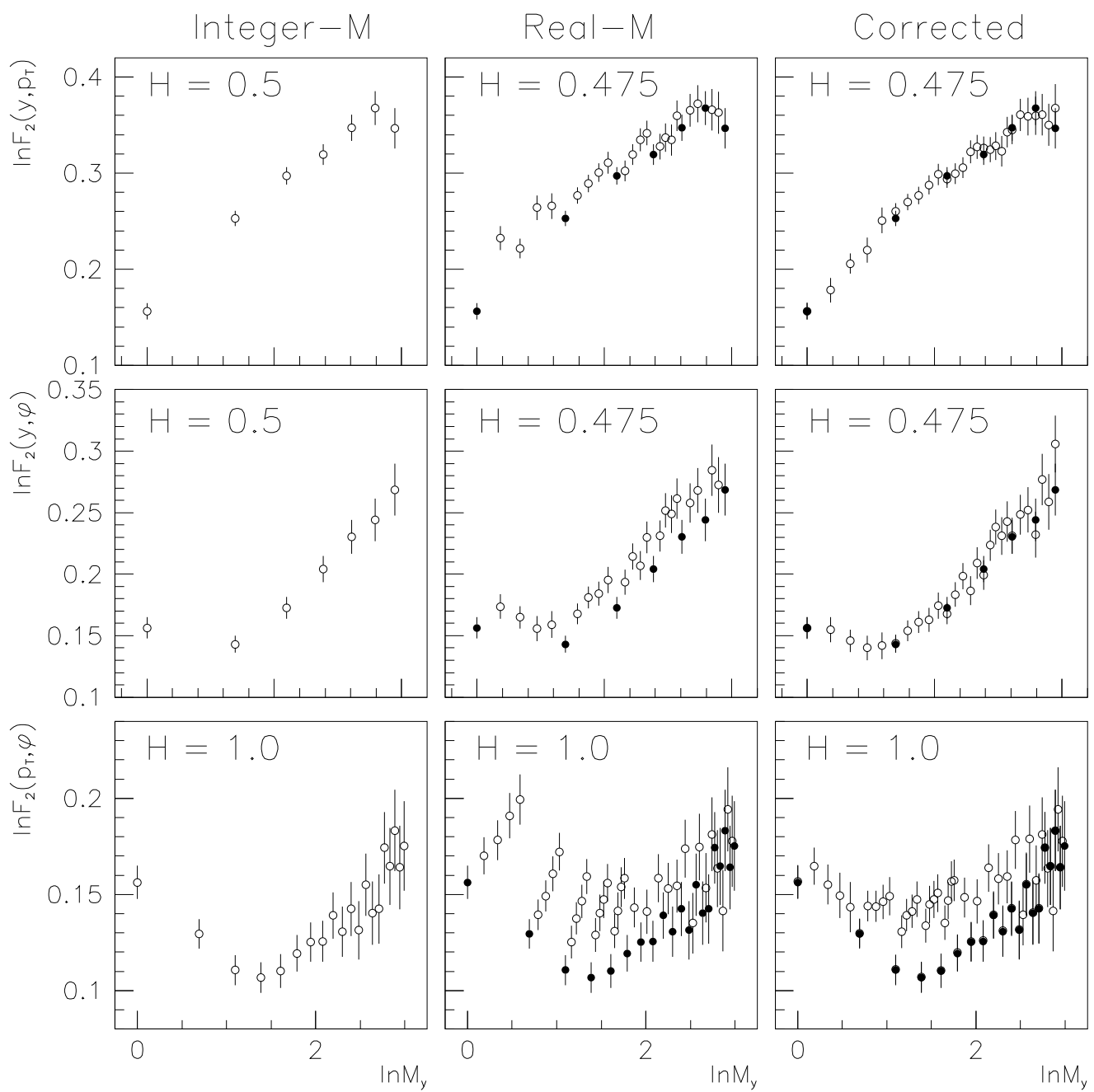
Fig.~3 \ The same as Fig.2 for two-dimensional self-affine FM as a function 
of $\ln M_y$.

\noindent 
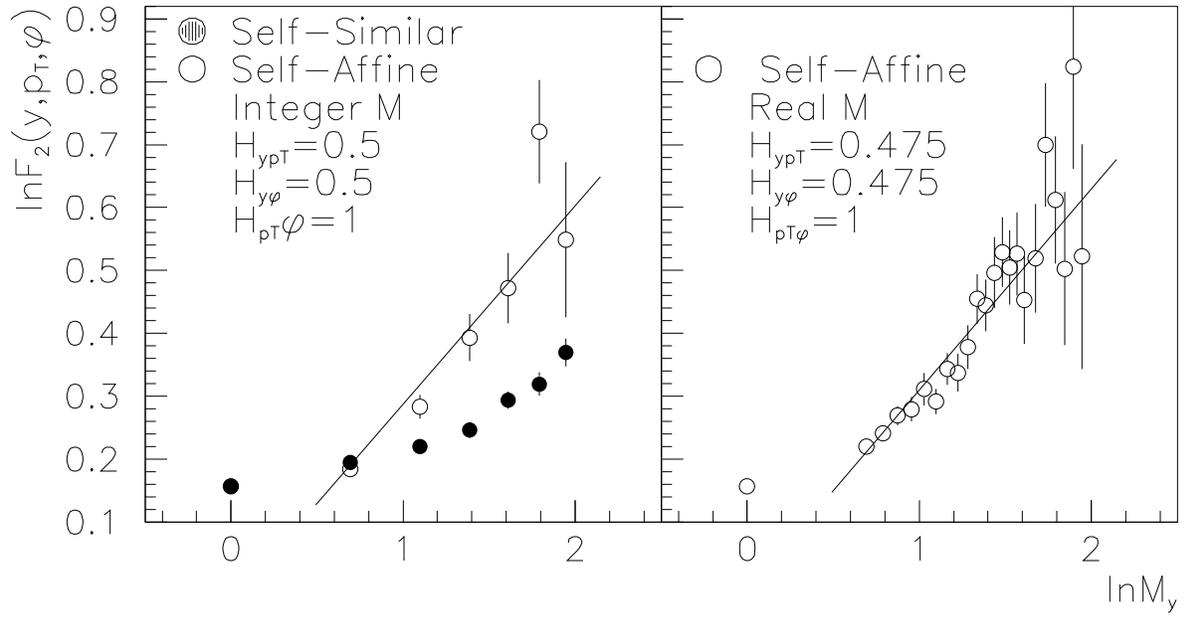
Fig.~4 \ The three-dimensional plots of ln$F_2$ versus ln$M$. The left 
figure gives the previous results for integer $M$. The right figure gives 
the corrected results of the self-affine analysis with real $M$ together
with a linear fit.

\end{document}